\documentclass{article}
\usepackage{amssymb}
\usepackage{amsfonts}
\usepackage{amsmath}

\input{tcilatex}

\begin{document}

\title{Towards a self-consistent model of analogue gravity}
\author{Andrzej Okninski \\
Physics Division, Politechnika Swietokrzyska, \\
Al. 1000-lecia PP 7, 25-314 Kielce, Poland}
\maketitle

\begin{abstract}
A nonlinear scalar field theory from which an effective metric can be
deduced is considered. This metric is shown to be compatible with
requirements of general relativity. It is demonstrated that there is a class
of solutions which fulfill both the nonlinear field equation as the Einstein
equations for this metric.
\end{abstract}

\section{Introduction}

There is increasing interest in building analogue models for general
relativity. The first analogue model was suggested for black holes and for
simulating Hawking evaporation \cite{Unruh1981}. Models, typically but not
always, based on condensed matter physics have been constructed to simulate
certain aspects of general relativity, see \cite{Barcelo2005} for very
extensive review. Existence of a general pattern has been proposed: \textit{%
That the occurrence of something like an approximate Lorentz symmetry, and
something like an approximate non-trivial "effective metric" might be an
inescapable general consequence of classical and quantum field theories
viewed as dynamical systems} \cite{Barcelo2001}. This claim can be given
more precise formulation. Namely, it has been shown that linearization of
any Lagrangian-based dynamics leads to effective Lorentzian geometry that
governs the propagation of the fluctuations \cite{Barcelo2001}.

In this work we consider a nonlinear scalar field theory from which an
effective metric follows in a covariant fashion, due to the appropriate
conservation law. This metric is shown to be compatible with requirements of
general relativity. Therefore we analyse consequences of the Einstein
equations for this metric. It is demonstrated that there is a class of
solutions which fulfill both the nonlinear field equation as the Einstein
equations for this metric. This result shows that the model of analogue
gravity is self-consistent for this class of solutions. Finally, the
obtained results are analysed and aims for further work are outlined.

In what follows tensor indices are denoted with Greek letters, $\mu =0,1,2,3$
and we shall always sum over repeated indices. The metric tensor of the
Minkowski space is defined as $\eta _{\mu \nu }=diag\,\left(
1,-1,-1,-1\right) $, $\left[ \eta _{\mu \nu }\right] $ denotes the
corresponding matrix, $x^{\mu }=\left( x^{0},\ x^{i}\right) $, $i=1,2,3$, $%
\varphi _{,\mu }=\frac{\partial \varphi }{\partial x^{\mu }}$ and a system
of units in which $c=\hslash =1$ is used throughout.

\section{Nonlinear classical field as a model of analogue gravity\label{1}}

Let the Lagrangian density $\mathcal{L}$ depends on the invariant $I=\varphi
_{,\mu }\varphi ^{,\mu }$ only. The Euler equations:

\begin{equation}
\dfrac{\partial }{\partial x^{\mu }}\dfrac{\partial \mathcal{L}}{\partial
\varphi _{,\mu }}=\dfrac{\partial \mathcal{L}}{\partial \varphi },
\label{Euler}
\end{equation}%
can be written in the following form:%
\begin{equation}
\gamma ^{\mu \nu }\varphi _{,\mu \nu }=0,  \label{BI1}
\end{equation}%
where the coefficients depend on the field, $\gamma ^{\mu \nu }=\gamma ^{\mu
\nu }\left( \varphi \right) $. We would like to write Eqs.(\ref{BI1}) in
covariant form $\left( \sqrt{-g}\,g^{\mu \nu }\varphi _{,\mu }\right) _{,\nu
}=0$ with some effective (inverse) metric $g^{\mu \nu }$, depending again on
the field $\varphi $. It will be possible if the coefficients $\gamma ^{\mu
\nu }$ are proportional to the conserved energy-momentum tensor $T^{\mu \nu }
$,

\begin{equation}
T^{\mu \nu }=\dfrac{\partial \mathcal{L}}{\partial \varphi _{,\mu }}\varphi
^{,\nu }-\eta ^{\mu \nu }\mathcal{L},\quad \left( T^{\mu \nu }\right) _{,\nu
}=0,  \label{T1}
\end{equation}%
i.e. we demand that $\lambda \left( \varphi \right) \gamma ^{\mu \nu }\left(
\varphi \right) =T^{\mu \nu }$, where $\lambda \left( \varphi \right) $ is a
non-singular scalar function of the field $\varphi $. In this case we can
set $\sqrt{-g}\,g^{\mu \nu }\overset{df}{=}const\,T^{\mu \nu }$.

It can be shown that there is a unique Lagrangian with the demanded
property. Indeed, for

\begin{equation}
\mathcal{L}=\sqrt{1-b\varphi _{,\mu }\varphi ^{,\mu }},  \label{Lagrangian}
\end{equation}%
we obtain the Euler equations which can be written in form (\ref{BI1}) with:%
\begin{equation}
\gamma ^{\mu \nu }=\left( -1+bI\right) \eta ^{\mu \nu }-b\varphi ^{,\mu
}\varphi ^{,\nu }.  \label{gamma}
\end{equation}

On the other hand, the energy-momentum tensor is computed as%
\begin{equation}
T^{\mu \nu }=\frac{\gamma ^{\mu \nu }}{\mathcal{L}}.  \label{T2}
\end{equation}%
We thus set 
\begin{equation}
\sqrt{-g}g^{\mu \nu }\overset{df}{=}-T^{\mu \nu }\ ,  \label{def}
\end{equation}%
$g=\det \left[ g_{\mu \nu }\right] $, where $\left[ g_{\mu \nu }\right] =%
\left[ g^{\mu \nu }\right] ^{-1}$, and due to the conservation law,$\left(
T^{\mu \nu }\right) _{,\nu }=0$, we have:%
\begin{equation}
\left( \sqrt{-g}g^{\mu \nu }\right) _{,\nu }=0,  \label{harmonicity}
\end{equation}%
i.e. the harmonicity or de Donder condition is satisfied, and Eqs.(\ref{BI1}%
), (\ref{gamma}) can be written in covariant form:%
\begin{equation}
\left( \sqrt{-g}g^{\mu \nu }\varphi _{,\mu }\right) _{,\nu }=0.  \label{BI2}
\end{equation}

Finally, computing determinants in (\ref{def}) we obtain $g=\det \left(
-T^{\mu \nu }\right) =-\mathcal{L}^{2}$ (note that $\det \left[ g^{\mu \nu }%
\right] =g^{-1}$) to arrive at the formula for the inverse effective metric:%
\begin{equation}
g^{\mu \nu }=-\frac{\gamma ^{\mu \nu }}{\mathcal{L}^{2}}.  \label{g1}
\end{equation}%
The choice of minus sign in (\ref{def}) guarantees that $g^{\mu \nu
}\rightarrow \eta ^{\mu \nu }$ for $b\rightarrow 0$.

To compute the metric tensor $g_{\mu \nu }$ the matrix $\left[ g^{\mu \nu }%
\right] $ is inverted:

\begin{equation}
\left[ g_{\mu \nu }\right] =\left[ g^{\mu \nu }\right] ^{-1}=\left[ \eta
_{\mu \nu }-b\varphi _{,\mu }\varphi _{,\nu }\right] .  \label{g2}
\end{equation}

Let us investigate signature of the matrix $\left[ g_{\mu \nu }\right] $.
After computing eigenvalues of $\left[ g_{\mu \nu }\right] $ it is easily
established that the metric has signature $+---$ provided that the
inequality $1-b\left( \varphi _{,0}^{2}-\varphi _{,1}^{2}-\varphi
_{,2}^{2}-\varphi _{,3}^{2}\right) >0$ holds (note that in this case the
equation (\ref{BI1}), (\ref{gamma}) is hyperbolic and, moreover, the
Lagrangian (\ref{Lagrangian}) is real). Thus the metric (\ref{g2}) fulfills
the fundamental requirement of the general relativity since for $1-b\left(
\varphi _{,0}^{2}-\varphi _{,1}^{2}-\varphi _{,2}^{2}-\varphi
_{,3}^{2}\right) >0$ there exists a transformation which transforms locally $%
g_{\mu \nu }$\ into $\eta _{\mu \nu }$.

\section{Solutions of the nonlinear field equation}

Equations (\ref{BI1}), (\ref{gamma}) have travelling solutions: 
\begin{equation}
\varphi \left( x\right) =\Phi _{\pm }\left( \pm k_{0}x_{0}-\overrightarrow{k}%
\cdot \overrightarrow{x}\right) ,\quad \left( k^{\mu }k_{\mu }=0\right) ,
\label{travel1}
\end{equation}%
what can be verified directly. It is an open question whether more
complicated solutions exist. It is interesting that in the $1+1$ dimensional
case:

\begin{equation}
\gamma ^{\mu \nu }\varphi _{,\mu \nu }=0,  \label{BI2Da}
\end{equation}%
\begin{equation}
\gamma ^{00}=-1-b\varphi ^{,1}\varphi ^{,1},\ \gamma ^{11}=1-b\varphi
^{,0}\varphi ^{,0},\ \gamma ^{01}=\gamma ^{10}=-b\varphi ^{,0}\varphi ^{,1}\
,  \label{BI2Db}
\end{equation}%
soliton-type solutions exist \cite{Barabashev1966, Whitham1974} and are
related to space-time quantization \cite{Barabashev1966}. Let us note here
that Eqs.(\ref{BI2Da}), (\ref{BI2Db}) are equivalent to the two-dimensional
Born-Infeld equation \cite{Whitham1974} (observe that in \cite{Whitham1974}
the following convention for the derivatives $\varphi _{t}\equiv \varphi
_{,0}=\varphi ^{,0}$ , $\varphi _{x}\equiv \varphi _{,1}=-\varphi ^{,1}$ was
used).

\section{Einstein equations}

We have seen that the nonlinear field theory considered in Section \ref{1}\
induces the effective metric (\ref{g2}) which has signature characteristic
of general relativity. It is now natural to consider the Einstein equations
in empty space: 
\begin{equation}
R_{\mu \nu }\left( \varphi \right) =0,  \label{R1}
\end{equation}%
where $R_{\mu \nu }\left( \varphi \right) $ is the Riemann curvature tensor
computed from the metric tensor (\ref{g2}).

We shall first analyse the Einstein equations within weak field
approximation. For small $b$ the Einstein equations can be written as:

\begin{equation}
R_{\nu \rho }=\tfrac{1}{2}g^{\mu \sigma }\left( g_{\mu \sigma ,\nu \rho
}-g_{\nu \sigma ,\mu \rho }-g_{\mu \rho ,\nu \sigma }+g_{\nu \rho ,\mu
\sigma }\right) +O\left( b^{2}\right) =0,  \label{R2}
\end{equation}%
and thank to the harmonicity condition (\ref{harmonicity}), Eq.(\ref{R2})
can be cast into form \cite{Dirac1975}:

\begin{equation}
R_{\nu \rho }=\tfrac{1}{2}g^{\mu \sigma }g_{\nu \rho ,\mu \sigma }+O\left(
b^{2}\right) =0.  \label{R3}
\end{equation}

Therefore it follows that travelling waves:%
\begin{equation}
g_{\nu \rho }=g_{\nu \rho }\left( k_{\mu }x^{\mu }\right) ,\quad \left(
k_{\mu }k^{\mu }=0\right)  \label{travel2}
\end{equation}%
fulfill the Einstein equations in vacuum with accuracy of order $O\left(
b^{2}\right) $.\medskip

This approximate result prompts us to check if the form (\ref{travel2})
fulfills the Einstein equations (\ref{R1}) exactly. The metric tensor (\ref%
{g2}) for 
\begin{equation}
\varphi \left( x\right) =\Phi \left( k_{\mu }x^{\mu }\right) ,\quad \left(
k_{\mu }k^{\mu }=0\right)   \label{phi}
\end{equation}%
is equal:%
\begin{equation}
\tilde{g}_{\mu \nu }=\eta _{\mu \nu }+\Omega h_{\mu \nu },\ h_{\mu \nu
}=-k_{\mu }k_{\nu }.\ \ \left( \Omega \equiv b\Phi ^{\prime 2}\left( k_{\mu
}x^{\mu }\right) ,\ k_{\mu }k^{\mu }=0\right)   \label{g3}
\end{equation}%
Since in what follows we shall need $k^{\mu }$ we find out that $k^{\mu }=%
\tilde{g}^{\mu \nu }k_{\nu }=\eta ^{\mu \nu }k_{\nu }$.

We shall compute the Riemann curvature tensor from the metric tensor (\ref%
{g3}):%
\begin{equation}
\tilde{R}_{\mu \nu \rho \sigma }=\tfrac{1}{2}\left( \tilde{g}_{\mu \sigma
,\nu \rho }-\tilde{g}_{\nu \sigma ,\mu \rho }-\tilde{g}_{\mu \rho ,\nu
\sigma }+\tilde{g}_{\nu \rho ,\mu \sigma }\right) +\tilde{\Gamma}_{\beta \mu
\sigma }\tilde{\Gamma}_{\nu \rho }^{\beta }-\tilde{\Gamma}_{\beta \mu \rho }%
\tilde{\Gamma}_{\nu \sigma }^{\beta }\ ,  \label{curv1}
\end{equation}%
where the Christoffel symbols are computed as:%
\begin{equation}
\tilde{\Gamma}_{\mu \nu \sigma }=\tfrac{1}{2}\left( \tilde{g}_{\mu \nu
,\sigma }+\tilde{g}_{\mu \sigma ,\nu }-\tilde{g}_{\nu \sigma ,\mu }\right) =%
\tfrac{1}{2}\Omega ^{\prime }\left( h_{\mu \nu }k_{\sigma }+h_{\mu \sigma
}k_{\nu }-h_{\nu \sigma }k_{\mu }\right) =-\tfrac{1}{2}\Omega ^{\prime
}k_{\mu }k_{\nu }k_{\sigma }.  \label{Ch}
\end{equation}

It follows that both terms quadratic in Christoffel symbols in (\ref{curv1})
vanish due to the condition $k_{\mu }k^{\mu }=0$ and the form $\tfrac{1}{2}%
\left( \tilde{g}_{\mu \sigma ,\nu \rho }-\tilde{g}_{\nu \sigma ,\mu \rho }-%
\tilde{g}_{\mu \rho ,\nu \sigma }+\tilde{g}_{\nu \rho ,\mu \sigma }\right) $
is zero by symmetry since $\tilde{g}_{\mu \sigma ,\nu \rho }=-\Omega
^{\prime \prime }k_{\mu }k_{\sigma }k_{\nu }k_{\rho }$. Thus the Riemann
curvature tensor vanishes: 
\begin{equation}
\tilde{R}_{\mu \nu \rho \sigma }=0,  \label{curv2}
\end{equation}%
and so vanishes the Ricci tensor. 

In conclusion, travelling waves (\ref{travel1}) satisfy the equations (\ref%
{BI1}), (\ref{gamma}) as well as the Einstein equations in empty space, $%
\tilde{R}_{\nu \rho }=0$. We can make a stronger statement concerning the
metric. Since the Riemann curvature tensor vanishes for the metric (\ref{g3}%
) then this metric describes the flat Minkowski space-time. In other words,
there exists a transformation converting $\tilde{g}_{\mu \nu }$ into $\eta
_{\mu \nu }$.

\section{Summary and open problems}

We have considered the nonlinear scalar field theory with the Lagrangian
density (\ref{Lagrangian}). Conservation law of the energy-momentum tensor (%
\ref{T2}) permits to write down equation of motion in covariant form (\ref%
{BI2}) and deduce the effective metric (\ref{g2}). There is only one
Lagrangian density, depending on the invariant $I=\varphi _{,\mu }\varphi
^{,\mu }$, which leads to covariant form of equation of motion (\ref{BI2}).
Signature of this metric is compatible with requirements of general
relativity. It has been next demonstrated that there is a class of solutions
which fulfill both the nonlinear field equation as the Einstein equations
for this metric. This result shows that this model of analogue gravity is,
for this class of solutions, self-consistent.

There are several open problems for further analysis. Firstly, question of
existence of soliton-type solutions for Eqs.(\ref{BI1}), (\ref{gamma}) must
be answered. Such solutions exist in two-dimensional space-time \cite%
{Whitham1974,Barabashev1966} and lead to space-time quantization \cite%
{Barabashev1966}. Secondly, it should be checked if the soliton solutions of
the two dimensional theory, or hypothetical solutions in four dimensions,
fulfill the Einstein equations to preserve self-consistency of the model.
Finally, the problem of space-time quantization, related to soliton-type
solutions, must be addressed.

\end{document}